%% file: ue.tex
\documentclass[11pt, twocolumn]{article}

\usepackage{geometry}
\usepackage{siunitx}
\usepackage{hyperref}
\usepackage{graphicx}
\usepackage[ruled,vlined,linesnumbered,algo2e]{algorithm2e}
\usepackage{algorithm}
\usepackage{algorithmic}
\input{math_commands.tex}

\usepackage{times,amsmath,epsfig}
\usepackage{booktabs}
\usepackage{amsfonts}
\usepackage{footmisc}
\usepackage{multirow}
\usepackage{graphics}
\usepackage{breqn}
\usepackage{xcolor}
\makeatletter
\newcommand\footnoteref[1]{\protected@xdef\@thefnmark{\ref{#1}}\@footnotemark}
\makeatother
\usepackage{subfig}

\newcommand{\footremember}[2]{%
    \footnote{#2}
    \newcounter{#1}
    \setcounter{#1}{\value{footnote}}%
}
\newcommand{\footrecall}[1]{%
    \footnotemark[\value{#1}]%
}

\begin{document}

\title{Uncertainty Detection and Reduction in Neural Decoding of EEG Signals}

\author{Tiehang Duan\footremember{alley}{Facebook, Inc. Seattle, WA 98109, USA}, Zhenyi Wang\footremember{trailer}{SUNY Buffalo, Buffalo, NY 14260, USA}, Sheng Liu\footrecall{trailer},  Sargur N. Srihari\footrecall{trailer}, Hui Yang\footnote{Netflix, Inc. Seattle, WA 98008, USA}
}

\date{}
\markboth{Journal of \LaTeX\ Class Files, Vol. 14, No. 8, August 2015}
{Shell \MakeLowercase{\textit{et al.}}: Bare Demo of IEEEtran.cls for IEEE Journals}
\maketitle

\begin{abstract}
 EEG decoding systems based on deep neural networks have been widely used in decision making of brain computer interfaces (BCI). Their predictions, however, can be unreliable given the significant variance and noise in EEG signals. Previous works on EEG analysis mainly focus on the exploration of noise pattern in the source signal, while the uncertainty during the decoding process  is largely unexplored. Automatically detecting and reducing such decoding uncertainty is important for BCI motor imagery applications such as robotic arm control etc. In this work, we proposed an uncertainty estimation and reduction model (UNCER) to quantify and mitigate the uncertainty during the EEG decoding process. It utilized a combination of dropout oriented method and Bayesian neural network for uncertainty estimation to incorporate both the uncertainty in the input signal and the uncertainty in the model parameters. We further proposed a data augmentation based approach for uncertainty reduction. The model can be integrated into current widely used 
EEG neural decoders without change of architecture. We performed extensive experiments for uncertainty estimation and its reduction in both intra-subject EEG decoding and cross-subject EEG decoding on two public motor imagery datasets, where the proposed model achieves significant improvement both on the quality of estimated uncertainty and the effectiveness of uncertainty reduction. 
\end{abstract}

\section{Introduction} 
Brain computer interfaces (BCI) aim to control computers and robots by directly monitoring human brain activities \cite{review2019}. A common way to record such signals is to use the Electroencephalography (EEG) equipments, which have the advantage of being non-invasive, high temporal resolution and relatively low acquisition cost. Motor imagery EEG signal, which records the activity of human brain during user imagined movements, is currently being actively explored given their wide applicability for motion restoration of disabled people \cite{Tariq2018}, neurorehabilitation \cite{Cantillo2018} and gaming control \cite{Liao2012}. 

Decisions made by BCI systems are based on EEG signal. However, the EEG signals are volatile and have significant variance across different subjects and even across different sessions of the same subject. Such variance are from two sources: 1) the noise oriented from the electrodes and the recording equipment, 2) the erratic nature of brain activity. The first source mainly contributes to the homoscedastic data noise, and the heteroscedastic data noise is mostly caused by the second source. An additional challenge for applying BCI systems is the efforts involved in labeling data for a new user, which results in a limited number of data points and brings undesired uncertainty when applying previously trained model classifier to a new user.

Quantifying and reducing the uncertainties of EEG decoding caused by such variance is important for motor imagery BCI applications.  Knowledge on when and to what extent the model is unsure about its decision helps mitigate hazard movements in BCI controlled robotic systems. Please note that the estimated uncertainty can be very different from the predicted probability of an EEG decoder. 
Significant uncertainty can incur even when the predicted probability of the target class is much higher than the other classes. This can happen when the model receives out of domain input data, \textit{e.g.}, EEG signals from a new user.

The uncertainty in EEG decoding comes from two sources: 1) uncertainty in the input signals, which is known as data uncertainty or aleatoric uncertainty, indicates the noise in EEG electrodes and recording equipments, 2) model uncertainty or epistemic uncertainty, which origins from the uncertainty lying in the parameters of the model \cite{kendall2017}. There is a high level of model uncertainty when there is a distribution shift between the training data and testing data. This is often the case for EEG classification as the pattern of EEG signals differs across subjects. In addition, there is a high level of model uncertainty when the amount of training EEG data is limited due to the significant amount of manual effort involved in data annotation. 

\begin{figure*} 
    \centering
  \subfloat{%
        \includegraphics[width=1.0\linewidth]{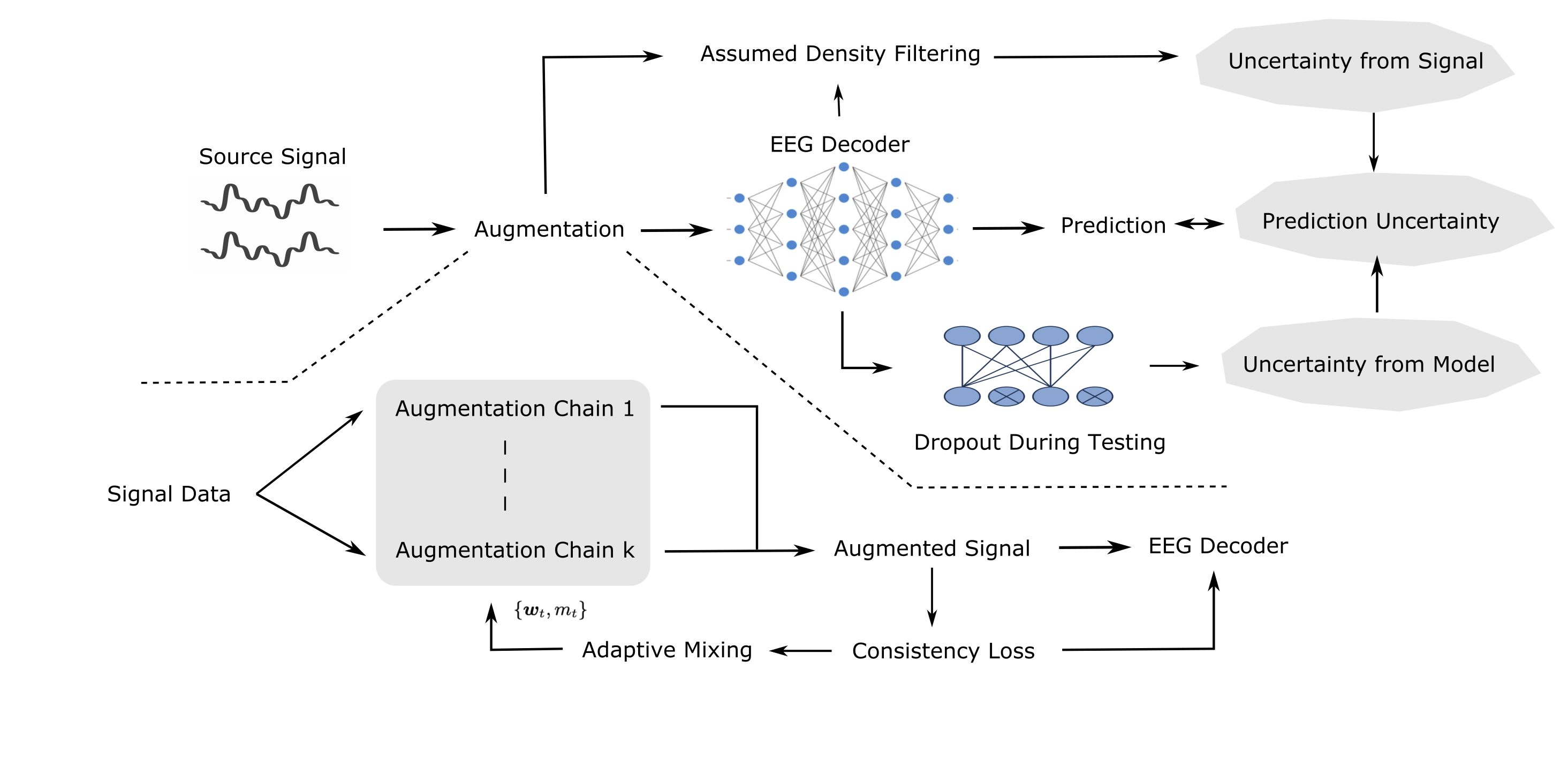}}
  \caption{Illustration on UNCER for uncertainty estimation and reduction in neural decoding of EEG signals}
  \label{fig8}
\end{figure*}

In this work, we proposed an \textbf{UNC}ertainty \textbf{E}stimation and \textbf{R}eduction framework (UNCER) for neural decoding of EEG signal with a focus on motor imagery BCI applications. The model quantifies data uncertainty through light weight probabilistic neural networks, which encode the variance in input data into probabilistic distribution, and forward propagates the moments all the way to output through Assumed Density Filtering (ADF). ADF is used to derive a probabilistic formulation of propagation for most of the operations in neural network such as convolution, pooling, ReLU etc. and is computationally effective in practice. For model uncertainty, UNCER performs estimation with drop-out similar to \cite{Gal2016Dropout}. Unlike direct modeling of model parameters with probability distributions, dropout based approaches don't need to change model architecture and also computationally efficient in nature. An uncertainty reduction approach is then proposed based on data augmentation, which performs a series of data augmentation operations dynamically on the fly. It adaptively augments the EEG signal by learning to mix the augmentation operations based on the data distribution, and works well for the distribution shift scenario in EEG data. 

In summary, our contributions are mainly three-fold:

1) We proposed a unified framework for uncertainty estimation and reduction in neural decoding of EEG signals. The proposed approach is computationally efficient and also versatile towards current neural EEG decoding models. It can be integrated into current widely used EEG decoders without changing the model architecture.

2) We perform interpretation analysis on the estimated uncertainty, and visualize the influence of individual channels towards the decoding uncertainty and compares it to the channel level
influence on the prediction. This intuitively
depicts the importance of individual channels towards output
prediction and uncertainty.

3) We performed extensive experiments for detailed comparison on the methods in terms of uncertainty, calibration and prediction accuracy etc. The result demonstrates the significant improvement brought by UNCER on the quality of estimated uncertainty and the effectiveness in uncertainty reduction. Calibration shows moderate gains and prediction accuracy is on par with comparison models.

\section{Related Work}

\subsection{Uncertainty in EEG Source Signal}
EEG signals are generated with electrodes monitoring brain activities such as rehearsal of a motor act for motor imagery tasks\cite{deng2012motor}. Proper decoding of its pattern allows it to be interpreted as control commands to devices such as wheelchair \cite{millan2004}\cite{chun2011motor}, robotic arms\cite{Onose2012} and gaming gadgets\cite{doud2011}. The EEG signal has high noise level with sources of noise including environment, recording equipment, experimental error etc., and its pattern is also highly subject specific\cite{Jiang2019}. Previous works have conducted research on noise estimation and variation analysis in the source signal\cite{maddirala2016}\cite{Goldenholz2009}\cite{Hassani2015}, with widely adopted noise estimation approaches include recursive least square filtering\cite{Noor_recursive}, discrete wavelet transform\cite{LiNieZhu2012}, and combination of adaptive filtering with discrete wavelet transformation etc.\cite{Madhale_adaptive} The noise estimation are performed in the pre-processing stage of EEG signal, and to our knowledge the uncertainty involved in the decoding process is largely unexplored.

\subsection{Neural Decoding of EEG Signal}

Neural decoding of EEG signal utilize different types of deep neural networks for EEG decoding and classification, e.g. \cite{Lawhern_2018} proposed a computationally efficient convolutional neural network (CNN) that is effective across different BCI platforms. \cite{Schirrmeister2017} adopted a novel cropped training strategy for the CNN model and achieved state of the art performance. \cite{ZhangD2018} come up with a cascade and parallel CNN architecture for improved performance. \cite{zhangd2019} utilized attentional mechanism on top of LSTM to effectively extract temporal features and achieved promising result. 
Compared to classic signal processing and machine learning approaches such as 
filter bank (FB) and common spatial pattern (CSP) \cite{kai2008}, power spectral density decomposition (PSD) \cite{Jatupaiboon2013} etc., most deep neural network oriented approaches are over-parameterized and produces deterministic results without uncertainty control in the process, which can leads to over-confident incorrect predictions and pose challenge for its deployment in real world EEG systems, with its unknown behavior towards factors such as generalization across different subjects, the fitting of confounders, and interpretability/explainability towards the result.

\subsection{Uncertainty Estimation in EEG Neural Decoding}
Uncertainty estimation is being actively explored in machine learning with important applications in fields such as auto driving and robotics \cite{ensemble2017}\cite{calibration2017}\cite{Loquercio2020AGF}. The estimation is done either through Bayesian probabilistic approaches \cite{lobato2015} or Monte-Carlo sampling based methods \cite{ensemble2017}, with the uncertainty encoded in the posterior distribution of model prediction. The Bayesian probabilistic approach model the parameter of neural network with probabilistic distributions, based on which uncertainties are analytically computed \cite{Wang2016Natural}. Simplification and approximation is needed to make the computation tractable in the process \cite{lobato2015}\cite{Wang2016Natural}. The approach require modification of the network optimization process, and additional efforts is needed to integrate them into existing deep neural network architectures. Probabilistic light weight neural network \cite{Gast2018} simplifies the previous models by adopting a partial probabilistic approach. The model requires minimal modification on existing networks, and proved effective for modeling uncertainty from data.

Monte-Carlo based approaches is another major branch of uncertainty estimation methods. Researchers found existing techniques such as dropout \cite{Gal2016Dropout} and ensemble models \cite{ensemble2017} could imply useful information on prediction uncertainty. Dropout base approaches \cite{Gal2016Dropout}\cite{concdrop} adopt dropout operation during test time for uncertainty estimation, which can be seen as modeling parameters in network to be Bernoulli distributed. Ensemble based approaches \cite{ensemble2017}\cite{Thurin_ensemble}\cite{Ashukha2020Pitfalls} estimates the posterior of model predictions by sampling on the differently trained neural networks. These Monte-Carlo based approaches effectively capture the uncertainty lies in model parameters. 

These models have been applied recently for uncertainty estimation in fields such as gravitational lensing of astrophysics \cite{Perreault2017}, biomarker quantification \cite{EatonRosen2018TowardsSD} and diagnosis of kidney injury \cite{Tomaev2019}.

\subsection{Uncertainty Reduction in EEG Neural Decoding}

Current exploration on uncertainty mainly focus on its quantization and few exploration are made in its reduction. Recently, researchers found data augmentation techniques turns out effective in this direction.  \cite{hendrycks*2020augmix} proposed Augmix and tested its effectiveness in terms of uncertainty reduction. \cite{Ayhan2018TesttimeDA} and \cite{Improve_calibration_augmentation} went on and explores different types of data augmentation for uncertainty improvement.

\section{Method}

We first introduce the functionality of uncertainty estimation for the proposed UNCER model, and then describe the uncertainty reduction process.

\subsection{Uncertainty Estimation of EEG Neural Decoders}

For accurate uncertainty estimation, the UNCER model considers both uncertainty from signal and uncertainty from model parameters by performing dropout operation on top of light weight probabilistic neural network. We first introduce the modeling of each type of uncertainty below, then we derive the composition of overall decoding uncertainty from the two sources. 

\subsubsection{Uncertainty propagated from input signal}
We take a probabilistic modeling approach of input uncertainty, and propagate the uncertainty through the neural decoder with assumed density filtering (ADF) \cite{boyen}\cite{Gast2018}. We denote the underlying noise free signal with $\mathbf{\bar{x}}$, and the noise corrupted input as $\mathbf{x}$, which can be modeled with distribution

\begin{equation} \label{eq5}
p(x|\bar{x})=\text{normal}(\bar{x}; u)
\end{equation}

where $u$ is the variance which depicts uncertainty in source signal.

The joint distribution of the hidden states in intermediate layers of decoder can be represented as

\begin{equation} \label{eq6}
p(z^{(1:l)})=\int_{x} p(x)\prod_{i=1}^{l}p(z^{i}|z^{i-1}) dx
\end{equation}

with $z^{0}=x$.

In assumed density filtering, the conditional probability is approximated as

\begin{equation} \label{eq7}
p(z^{i}| z^{i-1})=\text{normal} (\mu^{i}, v^{i})
\end{equation}

where the moments of the distribution are

\begin{equation} \label{eq8}
\mu^{i}=\mathbb{E}_{p(z^{i-1})}[f^{(i)}(z^{i-1})]
\end{equation}

\begin{equation} \label{eq9}
v^{i}=\mathbb{V}_{p(z^{i-1})}[f^{(i)}(z^{i-1})]
\end{equation}

where $f^{i}$ is the function performed with the $i$th layer of the neural network. $\mathbb{E}_{p(z^{i-1})}$ and $\mathbb{V}_{p(z^{i-1})}$ are the first and second moments of the underlying distribution. Forward propagating with this recursive conditional probabilistic rule produces $\mu^{l}$ and $v^{l}$ for the final decoding layer, which corresponds to the network prediction and its corresponding variance. 

The probabilistic propagation can be performed on common neural network structures such as convolutional layer, ReLU activation layer, batch normalization layer and pooling layer etc., corresponding to the different $f(z)$ functions. The propagation rule outlined in eq. \ref{eq8} and eq. \ref{eq9} can either be exactly derived or through probabilistic approximation. The convolutional layer performs linear operations in which $\mu^{i}$ and $v^{i}$ don't correlate with each other, allowing straight forward calculation of the two terms in closed form. The Relu layer, although nonlinear in nature, also leads to closed form solutions as derived in \cite{frey1999}. Other types of operations, e.g. max pooling,  requires probabilistic approximation for tractable computation, as detailed in \cite{jacobs2000}\cite{Jin2015RobustCN}. Please note this ADF propagation don't impose additional modification on the network structure. The number of parameters in the network also remains the same as its non-probabilistic counterpart. The only difference is that each layer receives the paired values (mean and variance) as input and also output the paired value to the next layer.

Different ways exist to estimate the noise level $u$ in the input signal. Previous works either uses a user defined constant \cite{Gal2016Dropout} or estimate the noise level from the data \cite{kendall2017}. Using a constant is computationally efficient, but in general it is not easy to accurately get the prior information of input noise characteristics. Learning the input noise from data is able to reflect the change from different input sources and increases the model's adaptation ability. However, its implementation requires tailored modification on the network architecture and hinders its application on existing classifiers. For applications in EEG signal analysis, the signal to noise ratio (SNR) is either available or can be readily computed with well developed techniques \cite{Goldenholz2009}\cite{Perez2018}. Another approach is to perform a grid search on a range of noise values for optimized NLL,
as NLL is minimized when the
magnitude of assumed input noise matches the underlying
ground truth.

\subsubsection{Uncertainty from model parameters} Given the randomness in the initialization and the training process of neural decoders, especially when amount of training data is limited, another source of uncertainty is brought in by the variation lying in model parameters. We model this source of uncertainty by performing dropout during testing, which is computationally efficient and easy to integrate into existing neural decoders. 

Dropout can be seen as the sampling process with parameters are approximated as Bernoulli distributed.  
Denoting parameter distribution after training as $P(W|X, Y)$, this approximation can be represented as

\begin{equation}
P(W|X, Y) = \text{Bernoulli}(W; \phi)
\end{equation}

where $\phi$ is the dropout rate. Binary variables are sampled out for each node in the network (except nodes of output layer). Each variable is 1 with probability $\phi$, corresponding to the nodes retained during the dropout. This process is performed during testing to estimate the resulting uncertainty on model output $y^*$. 

\begin{equation} \label{eq10}
v_{m}=\mathbb{E}_{p(y^{*}|x^{*})}(y^{*})^2-(\mathbb{E}_{p(y^{*}|x^{*})}(y^{*}))^2
\end{equation}

where $(x^{*}, y^{*})$ is testing data and

\begin{equation} \label{eq11}
 \mathbb{E}_{p(y^{*}|x^{*})}(y^{*})^2=\frac{1}{T}\sum_{t=1}^{T} y^{*}(x^{*}, W^{t})^2
\end{equation}

\begin{equation} \label{eq12}
 \mathbb{E}_{p(y^{*}|x^{*})}(y^{*})=\frac{1}{T}\sum_{t=1}^{T} y^{*}(x^{*}, W^{t})
\end{equation}

$T$ is the number of stochastic dropout forward passes for each test data point. The optimal dropout rate $\phi$ minimizes the KL divergence between the approximated Bernoulli distribution and the underlying parameter distribution, previous work \cite{Loquercio2020AGF} has shown this is equivalent to maximizing the log likelihood when input and hidden states are normal distributed. We perform grid search for optimal value of $\phi$ in the range of $[0, 1]$. 

\subsubsection{Estimating total uncertainty}

The variance from model parameter becomes independent of variance from source signal with post-training parameter distribution $p(W|X,Y)$ modeled as Bernoulli distributed $q(W; \phi)$, and we have


\begin{equation} \label{eq13}
\begin{split}
v_{p(y|x)}(y) & =\mathbb{E}_{p(y|x)}(yy^{T})-\mathbb{E}_{p(y|x)}(y)\mathbb{E}_{p(y|x)}(y)^T \\ &=\frac{1}{N}\sum_{n=1}^{N}v_{n}+\frac{1}{N}\sum_{n=1}^{N}\mu_{n}^{2}-\bar{\mu}^{2} \\ &=\frac{1}{N}\sum_{n=1}^{N}v_{n}+\frac{1}{N}\sum_{n=1}^{N}(\mu_{n}-\bar{\mu})^2 \\ & =v_{p(y|x)}^{d}(y) +v_{p(y|x)}^{m}(y) 
\end{split}
\end{equation}

where $N$ is the number of samples during test, $\mu_{n}$ and $v_{n}$ are from the ADF output, and $\bar{\mu}=\frac{1}{N}\sum_{n=1}^{N}\mu_{n}$. 

The overall workflow for estimating output uncertainty can be summarized into three steps:

(1) Train the classifier on EEG signals the same way as normal EEG neural decoders.

(2) Change the network into ADF propagation setting during test, and collect $N$ samples $\{\mu_{i}, v_{i}\}_{i\in [1,..,N]}$ with dropout rate $\phi$. 

(3) Compute predicted uncertainty based on eq. \ref{eq13}.

\subsection{Uncertainty Reduction with Data Augmentation}

We proposed a data augmentation based approach for uncertainty reduction in EEG decoding. The idea is to increase the prediction robustness towards different unseen data corruptions encountered during testing. The proposed method dynamically determines on the series of data augmentation to be performed on the EEG data, and then enforce a consistent embedding for the EEG neural decoder across the diversified augmentation operations based on consistency loss. The diversified transformations generated from the dynamically mixed augmentations helps inducing robustness and reducing uncertainty in the event of distribution shift in EEG data. It is easy to implement and also brings little computational overhead, allowing its wide application in BCI systems. 
\subsubsection{Augmentations}
To help foster the robustness and reduce uncertainty towards unseen noise, corruptions and distribution shifts during testing, we divides the augmentation operations $\gO$ into pseudo seen operation and pseudo unseen operations, with the two sets non-overlapping with each other. The augmentation process is then formulated as a meta learning procedure, with inner-loop optimizing the performance of pseudo-seen augmentation operations and outer-loop generalize to unseen augmentation. The augmentation performed in both the inner and outer loop involves composition of augmentation operations or augmentation chains. Each augmentation chain is constructed by composing from 1 to $d$ randomly selected augmentation operations. 

\begin{algorithm}[t!]
\small
	\caption{Data Augmentation for Uncertainty Reduction in EEG Decoding}
	\label{alg:uncer}
	\begin{algorithmic}[1]
		\STATE{\bf REQUIRE: Augmentation $\gO = \{osterize, rotate, \cdots\}$}; augmentation width $w$; augmentation depth $d$; the number of training iterations for each batch $\gT_{k}$ is $N_k$; inner-loop optimization steps $J$; inner-loop learning rate $\alpha$; regularization weight $\lambda$; original data $\vx_{orig}$
		 \vspace{-0.0cm}
		 \FOR{$k = 1$ to $N$}
		   \FOR{$t = 1$ to $N_k$}
		 \STATE augmented EEG data initialized as $\vec{\vx}_{b} = 0$
		 \STATE randomly split the augmentation operations  $\gO$ into non-overlapping pseudo-seen $\gS_t$ and unseen  $\gU_t$ operations
		 \STATE $\vx_{b} = \vx_{orig}$
		 		 \STATE $ \{\vw_t, m_t\}, \vh_{t+1}, \vc_{t+1} = \mathrm{LSTM}_{\bm{\phi}_t}(\mI_t, \vh_t, \vc_t)$
		 	\FOR{$i = 1$ to $w$}
		 	    	\FOR{$l = 1$ to $d$}
	    	    \STATE randomly sample operation $op \in \gS_t$ 
	    	    \STATE $\vx_{b} = op(\vx_{b})$
	    	    \ENDFOR
	    	    \STATE $\vec{\vx}_{b} += \vw_t^i \cdot \vx_{b}$ 
                \ENDFOR
                \STATE $\vec{\vx}_{b} = m_t \vx_{orig} + (1-m_t) \vec{\vx}_{b}$;
                \FOR{$k = 1$ to $J$}
                 \STATE Update model parameters $\vtheta_t$ based on loss function $\mathop{\mathbb{E}}_{ op \in \gS_t} \gL(\vec{\vx}_{b}, y_b, \vtheta, \bm{\phi}) + \lambda JS(\vec{\vx}_{b}, \vx_{orig})$
                \ENDFOR
                            \STATE augment data with pseudo unseen operations $\gU_t$ to be $\widehat{\vx}_{b}$; 
                \STATE update LSTM parameters $\bm{\phi}_{t}$ based on meta loss $  \mathop{\mathbb{E}}_{op \in \gU_t} \gL(\widehat{\vx}_{b}, y_b, \vtheta_{J}, \bm{\phi}) 
  + \lambda JS(\widehat{\vx}_{b}, \vx_{orig})$
             \ENDFOR
            \ENDFOR
	\end{algorithmic}
\end{algorithm}

\subsubsection{Adaptive Mixing}
The multiple augmentation chains are mixed with adaptive weights.  Given the nonstationary nature of EEG data, the augmentation strategy is data dependent and need to be dynamically learned. We proposed a self adaptive mixing approach for the sequential augmented EEG data, with the mixing weights sequentially computed through LSTM

\begin{equation}
    \vo_{t+1}, \vh_{t+1}, \vc_{t+1} = \mathrm{LSTM}_{\bm{\phi}_t}(\mI_t, \vh_t, \vc_t),
\end{equation}

where $\bm{\phi}_t$ is the model parameters, $\vh_t$ and $\vc_t$ is the hidden state and cell state for each datapoint in the batch of LSTM at time $t$, $\mI_t$ are the embedding of the EEG data of current time step $t$; $\vo_{t+1}$ is the output of LSTM at time $t+1$ for the mixing parameters, i.e., $ \vo_{t+1} = \{\vw_t, m_t\}$, where $m_t$ is the mixing weight between the augmented data and original raw memory data and $\vw_t$ is the mixing weights for individual augmentation chains. Sigmoid operations are applied on $m_t$ and softmax on $\vw_t$ for them to be in [0, 1] range. Details on the procedure is provided in Algorithm \ref{alg:uncer}.

\subsubsection{Consistency in Augmented EEG Signal}
A consistency loss is formulated to enforce the EEG decoder to have consistency prediction towards the diversified augmented EEG data. We minimize the Jensen-Shannon divergence among the posterior
distributions of the original EEG data $x_{orig}$ and its augmented versions $\widehat{\vx}_{b}$. The loss function $JS(\widehat{\vx}_{b}, \vx_{orig})$ can be obtained by calculating

\begin{gather}
    p_{mean} = (p_{\vx_{orig}}, p_{\widehat{\vx}_{b}})/2 \\
    \begin{split}
   JS(\widehat{\vx}_{b}, \vx_{orig}) \!=\!  (\mathbb{KL}(p_{\vx_{orig}}|p_{mean}) \!+\! \mathbb{KL}(p_{\widehat{\vx}_{b}}|p_{mean}) )/2
   \end{split}
\end{gather}

where $\mathbb{KL}$ denotes the KL Divergence between two distributions. $p_{\vx_{orig}}, p_{\widehat{\vx}_{b}}$ are the predicted probability of the original data and its augmentations. $p_{\vx_{orig}} = f_{\vtheta_t}(\vx_{orig})$, $p_{\widehat{\vx}_{b}} =  f_{\vtheta_t}(\widehat{\vx}_{b})$.

\section{Experiments}

\subsection{Dataset}

\begin{figure*} 
    \centering
  \subfloat[Left hand, Prediction \label{1b}]{%
        \includegraphics[width=0.25\linewidth]{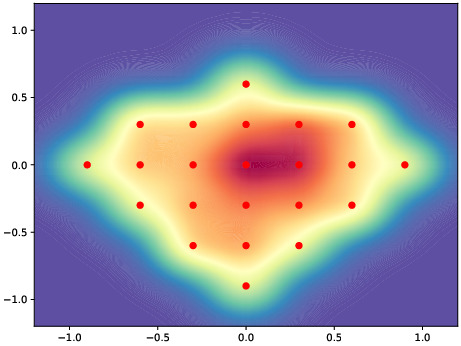}}
    \hfill
  \subfloat[Right hand, Prediction\label{1d}]{%
        \includegraphics[width=0.25\linewidth]{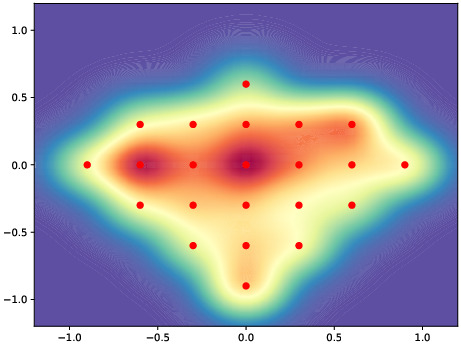}}
    \hfill
  \subfloat[Foot, Prediction\label{1d}]{%
        \includegraphics[width=0.25\linewidth]{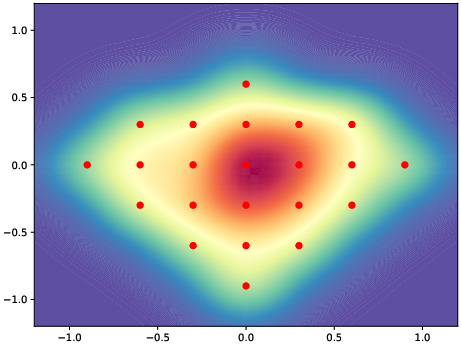}}
    \hfill
  \subfloat[Tongue, Prediction\label{1d}]{%
        \includegraphics[width=0.25\linewidth]{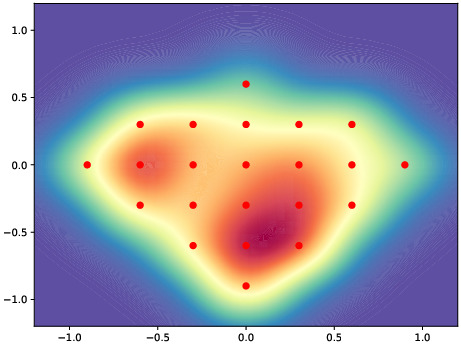}}
    \\
  \subfloat[Left hand, Uncertainty\label{1b}]{%
        \includegraphics[width=0.25\linewidth]{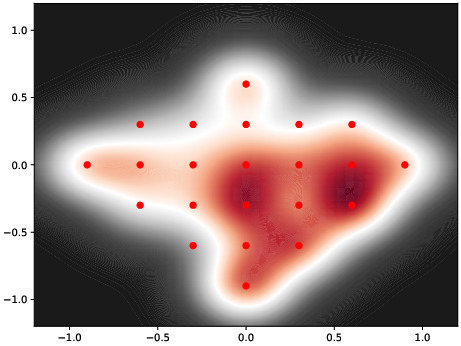}}
    \hfill
  \subfloat[Right hand, Uncertainty\label{1d}]{%
        \includegraphics[width=0.25\linewidth]{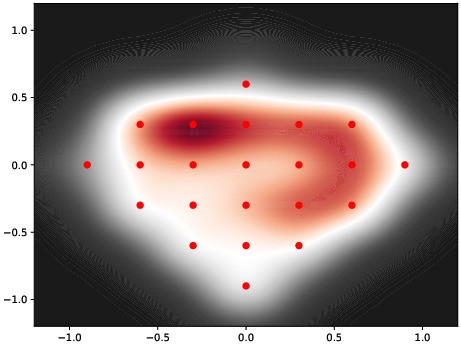}}
    \hfill
  \subfloat[Foot, Uncertainty\label{1d}]{%
        \includegraphics[width=0.25\linewidth]{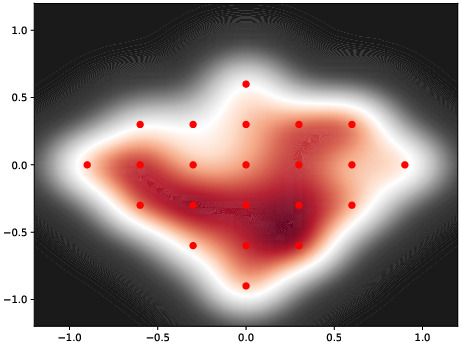}}
    \hfill
  \subfloat[Tongue, Uncertainty\label{1d}]{%
        \includegraphics[width=0.25\linewidth]{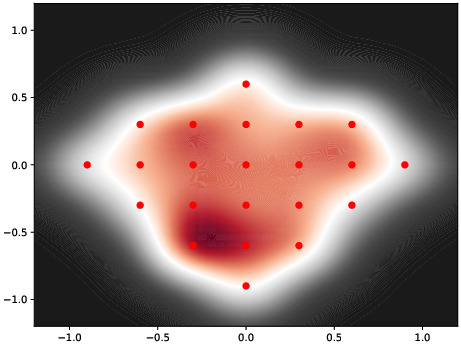}}
  \caption{Scalp topography heat maps visualizing the influence of individual channels on output prediction and uncertainty. The columns from left to right corresponds to left hand, right hand, foot and tongue. Top row shows the influence on prediction and bottom row reveals the influence on uncertainty. Given the dynamic and volatile nature of EEG signals, the result is averaged across 20 runs.}
  \label{fig8} 
\end{figure*}

We used two public BCI datasets in our experiment, which are BCI competition IV dataset 2a (abbreviated as BCI IV-2a below) \cite{Tangermann2012} \footnote{\url{http://bnci-horizon-2020.eu/database/data-sets}} and high gamma dataset \cite{Schirrmeister2017} \footnote{\url{https://github.com/robintibor/high-gamma-dataset}}. BCI IV-2a is relatively small and high gamma dataset contains a larger number of trials per subject. Details for each dataset is introduced below.

BCI IV-2a involves 9 subjects performing 4 class motor imaginary tasks. Each subject taking part in two sessions and each session consists of 288 trials. The tasks includes movement of left hand, right hand, both feet and tongue.  Signals are recorded with 22 electrodes at 250Hz sampling rate. A training phase and an evaluation phase were recorded on different days for each subject.

\begin{table}[htbp]
  \centering
  \caption{Comparison of different uncertainty estimation methods for intra subject classification on BCI-IV 2a dataset. The reported results are averaged across 10 runs. The highest performance are bolded and the runner up method are marked with $^{\dagger}$. Quality of estimated variance are measured with NLL, on which the proposed model outperformed comparison methods by at least 28\%. The model achieved either better or comparable result compared to other models in terms of calibration and prediction performance.}
  \resizebox{\columnwidth}{!}{%
    \begin{tabular}{l|c|cc|cc}
    \toprule
    Method & \multicolumn{1}{l|}{Variance Estimation} & \multicolumn{2}{c|}{Calibration} & \multicolumn{2}{c}{Performance} \\
    \midrule
          & \multicolumn{1}{c|}{NLL($\downarrow$)} & \multicolumn{1}{l}{Brier($\downarrow$)} & \multicolumn{1}{l|}{ECE($\downarrow$)} & \multicolumn{1}{l}{Acc.($\uparrow$)} & \multicolumn{1}{l}{ROC-AUC($\uparrow$)} \\
    \midrule
    Ensemble\cite{ensemble2017} & 12.4 & 0.108 & 0.276 & \textbf{0.705} & \textbf{0.832} \\
    BayesNet\cite{kendall2017}  & 32.4  & 0.117 & 0.295 & 0.681 & 0.813 \\
    Dropout\cite{Gal2016Dropout} & \hspace{2mm}7.45$^{\dagger}$  & \hspace{1mm}0.105$^{\dagger}$ & \hspace{1mm}0.271$^{\dagger}$ & 0.687 & 0.817 \\
    UNCER (ours)  & \textbf{5.39}  & \textbf{0.092} & \textbf{0.268} & \hspace{1mm}0.692$^{\dagger}$ & \hspace{1mm}0.824$^{\dagger}$ \\
    \bottomrule
    \end{tabular}%
    }
  \label{t1}%
\end{table}%

High gamma dataset is originally recorded with 128 electrodes, and in our experiment, we used the 44 channels covering the motor cortex region, in accordance with \cite{Schirrmeister2017}. 880 trials are performed on 14 subjects of balanced gender, with each trial consists of 4 seconds of recording on 4 classes of movement: left hand, right hand, both feet and rest. The signal is recorded with a BCI2000 device and then downsampled to 250 Hz, which is the same as BCI IV-2a dataset and allows the same hyper parameter setting for both datasets.

\subsection{Settings}

\begin{table}[htbp]
  \centering
  \caption{Comparison of different uncertainty estimation methods for intra subject classification on high gamma dataset. Other experiment settings are kept the same as table \ref{t1}}
    \resizebox{\columnwidth}{!}{%
    \begin{tabular}{l|c|cc|cc}
    \toprule
    Method & \multicolumn{1}{l|}{Variance Estimation} & \multicolumn{2}{c|}{Calibration} & \multicolumn{2}{c}{Performance} \\
    \midrule
          & \multicolumn{1}{c|}{NLL($\downarrow$)} & \multicolumn{1}{l}{Brier($\downarrow$)} & \multicolumn{1}{l|}{ECE($\downarrow$)} & \multicolumn{1}{l}{Acc.($\uparrow$)} & \multicolumn{1}{l}{ROC-AUC($\uparrow$)} \\
    \midrule
    Ensemble\cite{ensemble2017} & \hspace{1mm}1.16$^{\dagger}$ & 0.069 & 0.205 & \hspace{1mm}0.873$^{\dagger}$ & \hspace{1mm}0.922$^{\dagger}$\\
    BayesNet\cite{kendall2017}  & 2.75  & 0.074 & 0.217 & 0.865 & 0.908 \\
    Dropout\cite{Gal2016Dropout} & 1.48  & \hspace{1mm}0.062$^{\dagger}$ & \hspace{1mm}0.179$^{\dagger}$ & 0.871 & 0.919 \\
    UNCER (ours)  & \textbf{0.64}  & \textbf{0.053} & \textbf{0.144} & \textbf{0.881} & \textbf{0.927} \\
    \bottomrule
    \end{tabular}%
    }
  \label{t2}%
\end{table}%

We evaluated our uncertainty estimation model for both intra subject classification and cross subject classification. For intra subject classification, recording from the same subject are split between training and testing, and for the cross subject setting, we leave one subject out each time and model is trained on the remaining subjects. For recording of each trial, we break it down to segments with window size of 400, and a stride size of 50 between adjacent segments. The models are implemented with Pytorch and runs on a single TITAN-V GPU. The models are fully trained for 40 epochs before evaluation, and we used Adam optimizer with learning rate set to 0.001. We performed dropout during testing to estimate the model uncertainty, and the dropout rate $\phi =0.1$ is determined by grid search on 40 log range uniformly distributed rates lying between $[0, 1]$.

\begin{table}[htbp]
  \centering
  \caption{Comparison of different uncertainty estimation methods for cross subject classification on BCI-IV 2a dataset. Other experiment settings are kept the same as table \ref{t1}}
    \resizebox{\columnwidth}{!}{%
    \begin{tabular}{l|c|cc|cc}
    \toprule
    Method & \multicolumn{1}{l|}{Variance Estimation} & \multicolumn{2}{c|}{Calibration} & \multicolumn{2}{c}{Performance} \\
    \midrule
          & \multicolumn{1}{c|}{NLL($\downarrow$)} & \multicolumn{1}{l}{Brier($\downarrow$)} & \multicolumn{1}{l|}{ECE($\downarrow$)} & \multicolumn{1}{l}{Acc.($\uparrow$)} & \multicolumn{1}{l}{ROC-AUC($\uparrow$)} \\
    \midrule
     Ensemble\cite{ensemble2017}  &    76.8  &    \hspace{1mm}0.158$^{\dagger}$  &    0.340  &  \textbf{0.523}  &   \textbf{0.715}  \\
     BayesNet\cite{kendall2017} &        130.1  &    0.162  &      0.343  &        0.487  &        0.692  \\
     Dropout\cite{Gal2016Dropout}  &     \hspace{1mm}68.7$^{\dagger}$  &      \textbf{0.153}  &     \hspace{1mm}0.331$^{\dagger}$  &     0.494  &             0.688  \\
     UNCER (ours)  &           \textbf{46.5}  &    \hspace{1mm}0.158$^{\dagger}$  &     \textbf{0.317}  &      \hspace{1mm}0.506$^{\dagger}$  &             \hspace{1mm}0.695$^{\dagger}$  \\
    \bottomrule
    \end{tabular}%
    }
  \label{t3}%
\end{table}%

\begin{table}[htbp]
  \centering
  \caption{Comparison of different uncertainty estimation methods for cross subject classification on high gamma dataset. Other experiment settings are kept the same as table \ref{t1}}
    \resizebox{\columnwidth}{!}{%
    \begin{tabular}{l|c|cc|cc}
    \toprule
    Method & \multicolumn{1}{l|}{Variance Estimation} & \multicolumn{2}{c|}{Calibration} & \multicolumn{2}{c}{Performance} \\
    \midrule
          & \multicolumn{1}{c|}{NLL($\downarrow$)} & \multicolumn{1}{l}{Brier($\downarrow$)} & \multicolumn{1}{l|}{ECE($\downarrow$)} & \multicolumn{1}{l}{Acc.($\uparrow$)} & \multicolumn{1}{l}{ROC-AUC($\uparrow$)} \\
    \midrule
    Ensemble\cite{ensemble2017} & \hspace{1mm}1.66$^{\dagger}$  & 0.102 & 0.307 & 0.771 & \hspace{1mm}0.885$^{\dagger}$ \\
    BayesNet\cite{kendall2017} & 2.87  & 0.107 & 0.284 & 0.768 & 0.879 \\
    Dropout\cite{Gal2016Dropout} & 1.92  & \hspace{1mm}0.095$^{\dagger}$ & \hspace{1mm}0.267$^{\dagger}$ & \hspace{1mm}0.774$^{\dagger}$ & 0.883 \\
    UNCER (ours) & \textbf{1.24}  & \textbf{0.086} & \textbf{0.251} & \textbf{0.796} & \textbf{0.894} \\
    \bottomrule
    \end{tabular}%
    }
  \label{t4}%
\end{table}%

The EEG neural decoder used in our experiment is a 3 layer CNN network similar to EEGNet \cite{Lawhern_2018}. 
The decoder involves different types of convolution layers, with first layer consists temporal convolution filters to learn frequency information, followed by batch normalization. The second layer involves depthwise convolutions with temporal specific spatial filters. Zero padding is done to maintain the original data dimension, after which batch normalization and dropout are applied. The third layer performs pointwise convolution. Its post processing is the same as the second layer. The prediction layer is a single fully connected layer followed by softmax operation.

\begin{table*}[htbp]
  \centering
  \caption{Influence of number of samples collected in dropout on NLL, the estimated NLL converges with 200 samples for both BCI-IV 2a and high gamma dataset.}
    \begin{tabular}{l|rrrrrrrr}
    \toprule
    Num Samples & 25    & 50    & 75    & 100   & 125   & 150   & 175   & 200 \\
    \midrule
    Scenarios & \multicolumn{8}{c}{NLL}
    \\
    \midrule
    BCI-IV 2a intra & 174.7 & 75.9  & 27.4  & 7.09  & 5.97  & 5.74  & 5.66  & 5.39 \\
    BCI-IV 2a cross & 2586.3 & 583.2 & 211.8 & 124.3 & 109.5 & 74.6  & 51.9  & 46.5 \\
    High gamma intra & 323.2 & 24.6  & 2.71  & 1.86  & 1.22  & 1.03  & 0.71  & 0.64 \\
    High gamma cross & 856.8 & 71.5  & 6.83  & 2.72  & 1.91  & 1.58  & 1.39  & 1.24 \\
    \bottomrule
    \end{tabular}%
  \label{t9}%
\end{table*}%

For uncertainty reduction, a total of 8 different types of corruptions are applied, with the details on each type of corruption can be find in Table \ref{corruption-types}. In each iteration, we randomly select 6 of them as seen operations and the other 2 as unseen operations.

\begin{table*}[htbp]
  \centering
  \caption{Comparison of different data augmentation methods for uncertainty reduction on BCI-IV 2a dataset.}
    \begin{tabular}{c|l|c|cc|ccc}
    \toprule
    \multirow{2}[4]{*}{Scenario} & \multicolumn{1}{c|}{\multirow{2}[4]{*}{Method}} & Variance Estimation & \multicolumn{2}{c|}{Calibration} & \multicolumn{3}{c}{Performance} \\
\cmidrule{3-8}          &       & NLL   & Brier & ECE   & Acc.  & ROC-AUC & Corr. Err \\
    \midrule
    \multirow{4}[2]{*}{Intra-Subject} & MixUp & 8.12  & 0.102 & 0.268 & 0.676 & 0.802 & 0.564 \\
          & MaxUp & 7.21  & 0.106 & 0.265 & 0.669 & 0.781 & 0.559 \\
          & Augmix & 4.83  & 0.087 & 0.254 & 0.695 & 0.818 & 0.482 \\
          & UNCER & \textbf{2.95}  & \textbf{0.078} & \textbf{0.231} & \textbf{0.712} & \textbf{0.83}  & \textbf{0.458} \\
    \midrule
    \multirow{4}[2]{*}{Cross-Subject} & MixUp & 55.4  & 0.161 & 0.336 & 0.479 & 0.673 & 0.677 \\
          & MaxUp & 48.9  & 0.153 & 0.319 & 0.473 & 0.66  & 0.683 \\
          & Augmix & 35.2  & 0.122 & 0.281 & 0.521 & 0.694 & 0.626 \\
          & UNCER & \textbf{20.7}  & \textbf{0.109} & \textbf{0.248} & \textbf{0.538} & \textbf{0.727} & \textbf{0.581} \\
    \bottomrule
    \end{tabular}%
  \label{bci-iv-reduction}%
\end{table*}%

\subsection{Metrics}

We evaluated the model in terms of three different aspects: the quality of estimated variance, calibration and performance. 

The quality of estimated variance is reflected by the Negative log likelihood (NLL) on the prediction, which is defined as $\text{NLL} = \frac{1}{2} log(v)+\frac{1}{2v}(y-\bar{y})^2$. It is a major metric for measuring uncertainty \cite{ensemble2017}\cite{Gal2016Dropout}. Higher values of NLL depicts lower confidence of capturing the ground truth based on predicted output, and vice versa.

Calibration metrics such as Brier score \cite{Bradley2008} and ECE score \cite{calibration2017} measures the confidence of model output for classification tasks and is reported here in complementary of NLL. Brier score computes the squared error between predicted logits and one hot label encoding, defined as $\text{BS}= K^{-1}\sum_{k=1}^{K}(t_{k}-p(y_{k}))^{2}$, where $K$ is the number of classes, $t_{k}$ are elements of one hot label encoding and $p(y_{k})$ are the predicted logits. ECE score is $L^{1} norm$ of the difference between predicted logits and accuracy. Please note the evaluation of Brier score and ECE score only depend on the predicted logits and don't involve the estimated variance. Both Brier score and ECE score are proper scoring rules as defined in \cite{Tilmann2007}. We reported the accuracy and AUC-ROC score in our experiments to evaluate on the model's prediction and study the relationship between the model's performance and its predicted uncertainty. We also evaluated on corruption error during uncertainty reduction analysis, which is the average of error rate across the different types of corruptions, to provide more information on model performance.

\subsection{Uncertainty Estimation Analysis}

We compared the UNCER model with several other models applicable to EEG uncertainty analysis. The method of Gal et al. \cite{Gal2016Dropout} is the dropout oriented uncertainty estimation approach. The model of Kendall et al. \cite{kendall2017} compute output uncertainty utilizing Bayesian neural network. The model proposed in \cite{ensemble2017} is a robust uncertainty estimation model based on ensemble method.

We evaluate the uncertainty estimation model for both intra subject classification and cross subject classification on the two public datasets, with BCI-IV 2a dataset being relatively small and high gamma dataset considerably larger. Cross subject classification is a more challenging task compared to intra subject classification given the significant variability lies in the EEG signal across different subjects.

Table \ref{t1} shows the result for intra subject classification on BCI-IV 2a dataset. The proposed model outperformed comparison methods by at least 28\% on NLL, which is the major metric to evaluate the quality of estimated variance. The proposed model achieved comparable performance on Brier score and ECE score compared to other models, which shows the model's calibration is not significantly influenced by the different uncertainty estimation approaches. Ensemble oriented method \cite{ensemble2017} achieved slightly better result on accuracy and ROC-AUC, which can be attributed to the joint decision making of the mixture of models. The result on high gamma dataset with intra subject setting is reported in table \ref{t2}, where the proposed model achieved a 44.8\% improvement in terms of NLL. The models are better calibrated 
on high gamma dataset compared to BCI-IV 2a dataset, as reflected from the Brier score and ECE score. This is attributed to the lower noise level in its signal and also its larger dataset size allowing more thorough training.

\begin{table*}[htbp]
  \centering
  \caption{Comparison of different data augmentation methods for uncertainty reduction on high gamma dataset.}
    \begin{tabular}{c|l|c|cc|ccc}
    \toprule
    \multirow{2}[4]{*}{Scenario} & \multicolumn{1}{c|}{\multirow{2}[4]{*}{Method}} & Variance Estimation & \multicolumn{2}{c|}{Calibration} & \multicolumn{3}{c}{Performance} \\
\cmidrule{3-8}          &       & NLL   & Brier & ECE   & Acc.  & ROC-AUC & Corr. Err \\
    \midrule
    \multirow{4}[2]{*}{Intra-Subject} & MixUp & 1.06  & 0.047 & 0.144 & 0.878 & 0.909 & 0.348 \\
          & MaxUp & 1.13  & 0.059 & 0.156 & 0.857 & 0.895 & 0.363 \\
          & Augmix & 0.6   & 0.043 & 0.129 & 0.886 & 0.924 & 0.259 \\
          & UNCER & \textbf{0.52}  & \textbf{0.038} & \textbf{0.121} & \textbf{0.892} & \textbf{0.931} & \textbf{0.235} \\
    \midrule
    \multirow{4}[2]{*}{Cross-Subject} & MixUp & 2.86  & 0.087 & 0.259 & 0.783 & 0.865 & 0.563 \\
          & MaxUp & 3.09  & 0.084 & 0.262 & 0.774 & 0.846 & 0.58 \\
          & Augmix & 2.34  & 0.077 & 0.245 & 0.791 & 0.872 & 0.462 \\
          & UNCER & \textbf{1.73}  & \textbf{0.073} & \textbf{0.239} & \textbf{0.804} & \textbf{0.883} & \textbf{0.417} \\
    \bottomrule
    \end{tabular}%
  \label{high-gamma-red}%
\end{table*}%

We further evaluated the model performance under cross subject classification settings, and the results on BCI-IV 2a and high gamma dataset are reported respectively in Table \ref{t3} and table \ref{t4}. The accuracy and ROC-AUC sees a significant drop compared to their intra subject counterparts, as cross subject classification is a more challenging setting compared to intra subject classification. We also observed calibration error 
increase with the adoption of a more challenging task setting. NLL sees an evident increase compared to intra subject settings with the longer tailed output distribution yields reduced likelihood. For both datasets, the proposed model performs consistently better than the comparison methods in terms of variance estimation while calibrating to a similar degree as the other comparison models.

We explore the influence of individual channels towards the output uncertainty and compares it to the channel level influence on the prediction. The influence is quantified by adopting the occlusion based interpretation approach introduced in \cite{Zeiler14visualizingand}, which is an end to end approach and don't impose modification on the model architecture. Each channel is blocked out individually and the influence of its absence towards the output prediction and uncertainty is recorded. A channel is considered important for the predictive uncertainty if the uncertainty significantly increased with the specific channel occluded from the input. This influence from individual channels are visualized in the form of heatmaps. Fig. \ref{fig8} shows example topography heat maps on prediction and uncertainty for each of the BCI-IV four motor imaginary tasks, namely left hand, right hand, foot and tongue. The heatmap intuitively depicts the importance of individual channels towards output prediction and uncertainty. To mitigate the randomness and volatility in brain signal, the result is achieved by averaging across 20 runs. Result shows output uncertainty is influenced by broader regions of human brain in general compared to output prediction.

We further explores on the influence of number of samples collected during dropout on NLL. The result is shown in Table \ref{t9} for each of the experiment settings. The estimated NLL tend to converge with 200
samples for both BCI-IV 2a and high gamma dataset.

\begin{figure} 
    \centering
  \subfloat[\label{7a}]{%
        \includegraphics[width=0.5\linewidth]{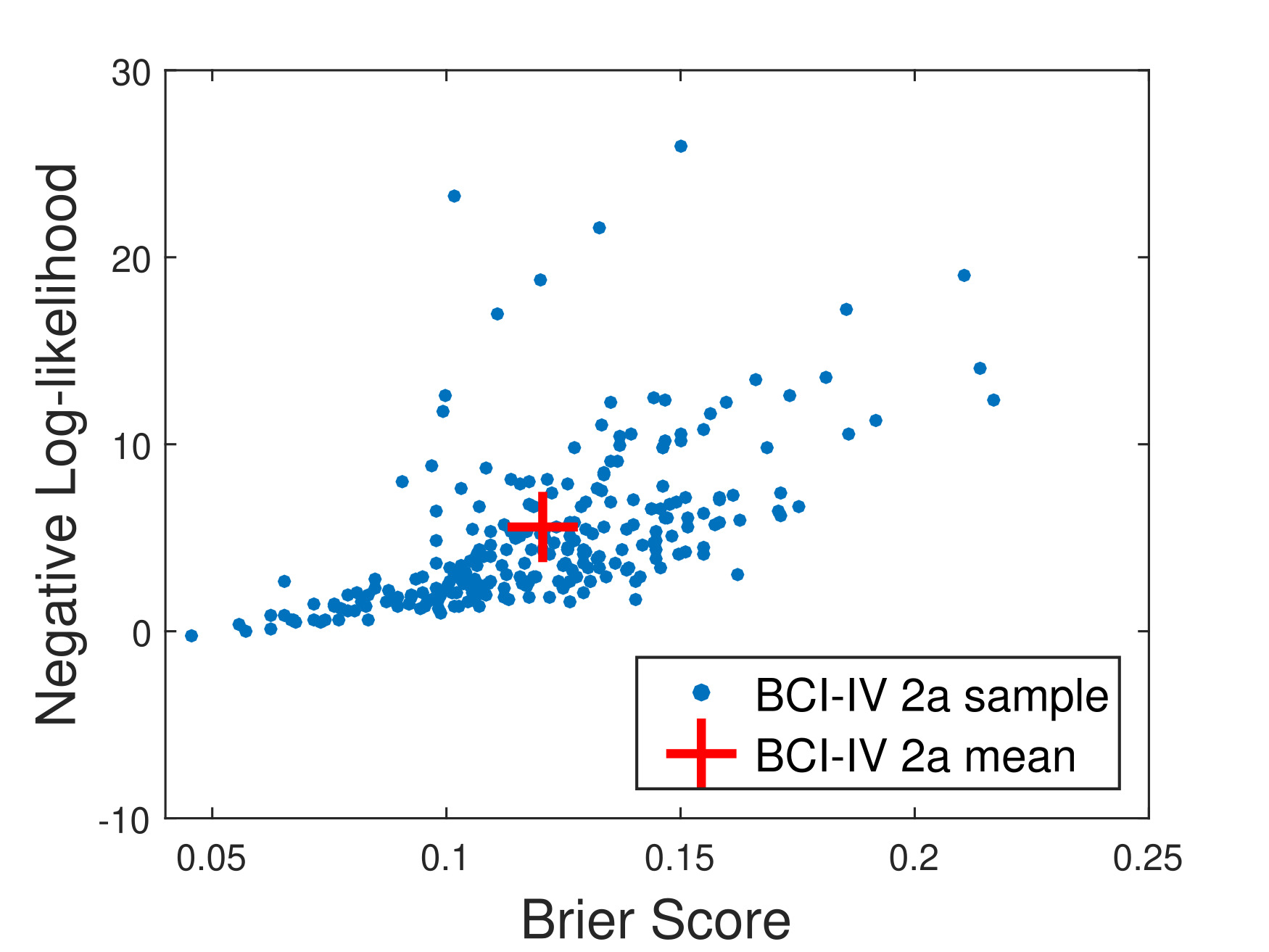}}
    \hfill
  \subfloat[\label{7b}]{%
        \includegraphics[width=0.5\linewidth]{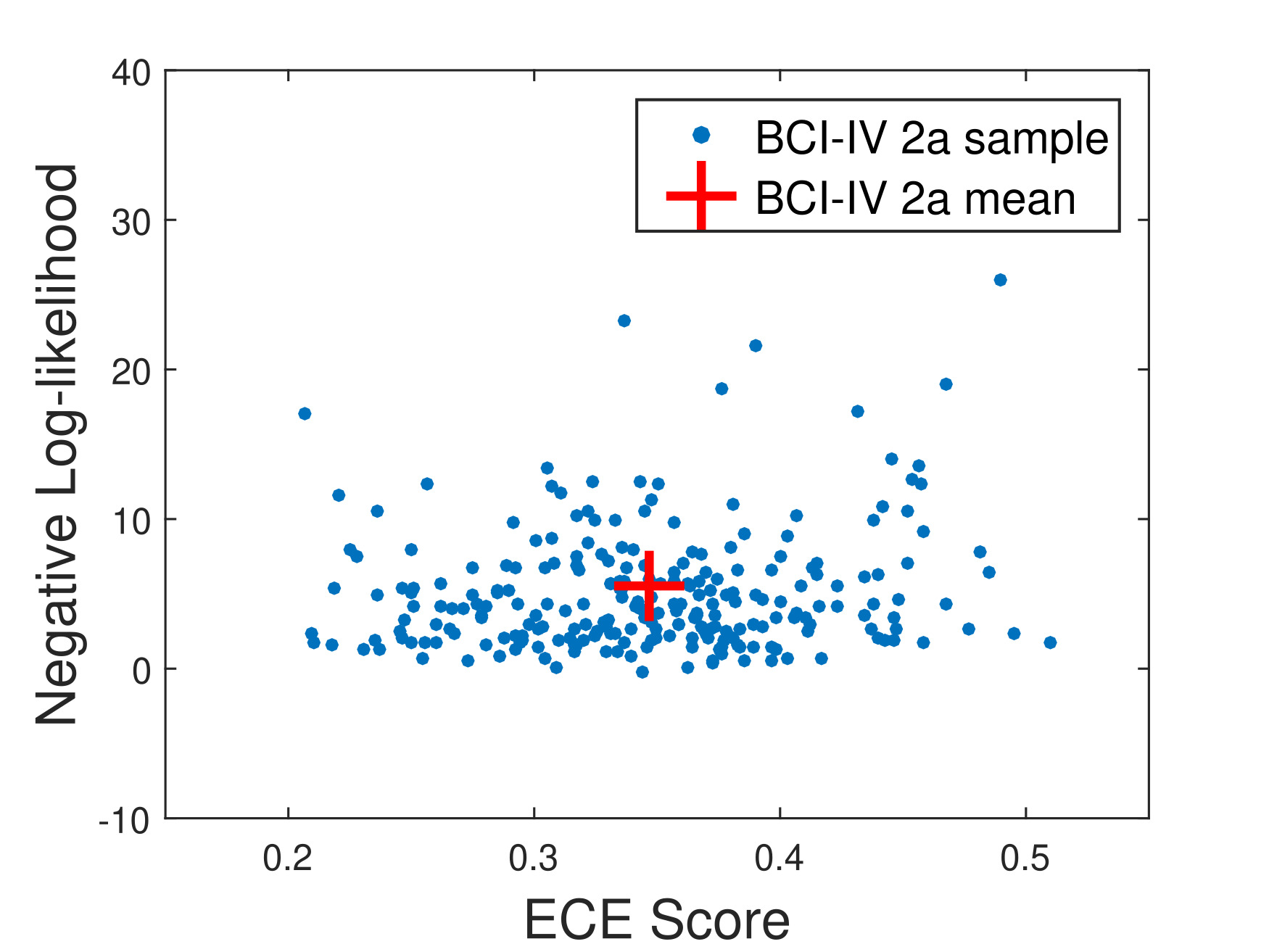}}
\\
  \subfloat[\label{7c}]{%
        \includegraphics[width=0.5\linewidth]{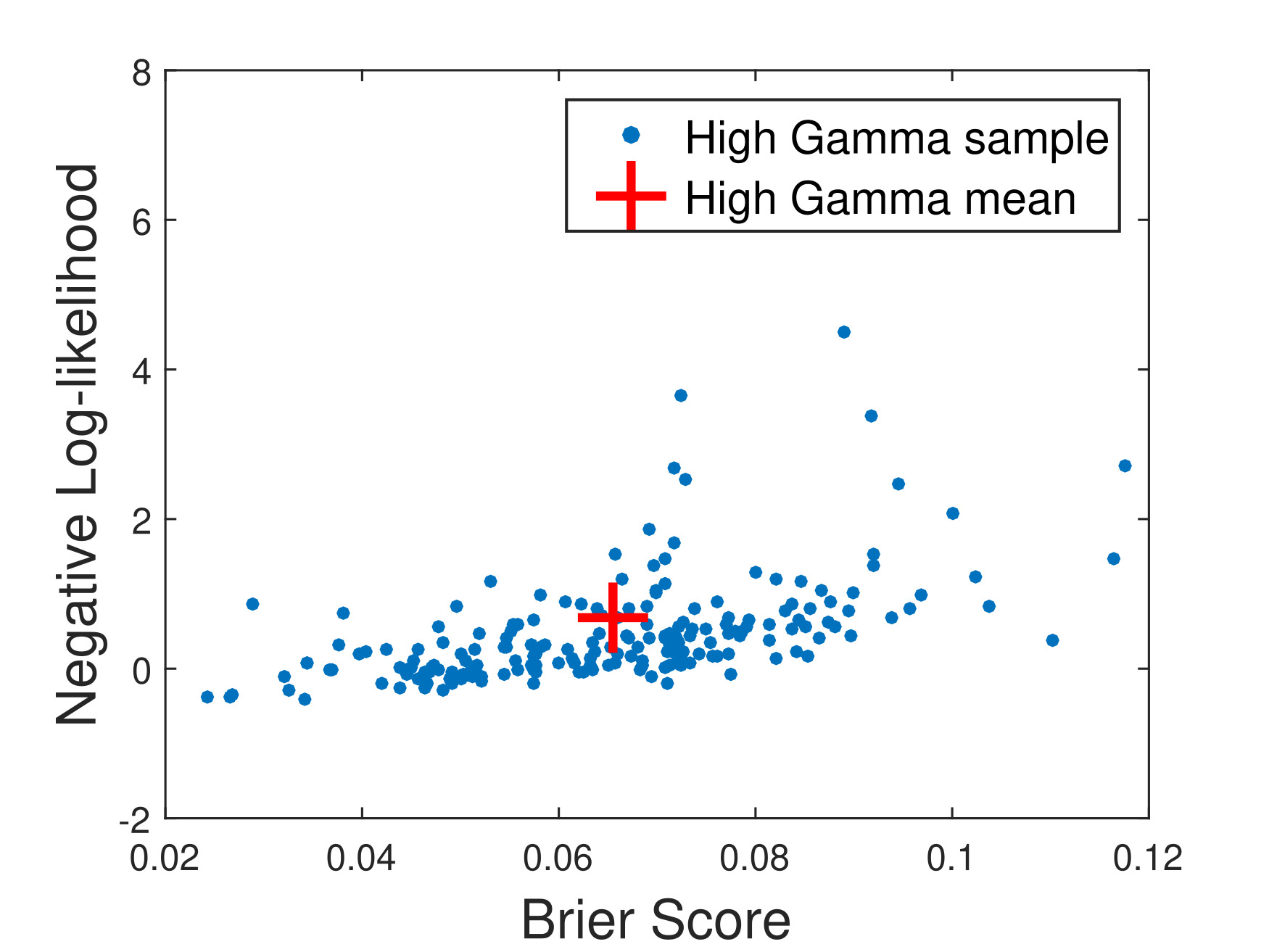}}
    \hfill
  \subfloat[\label{7d}]{%
        \includegraphics[width=0.5\linewidth]{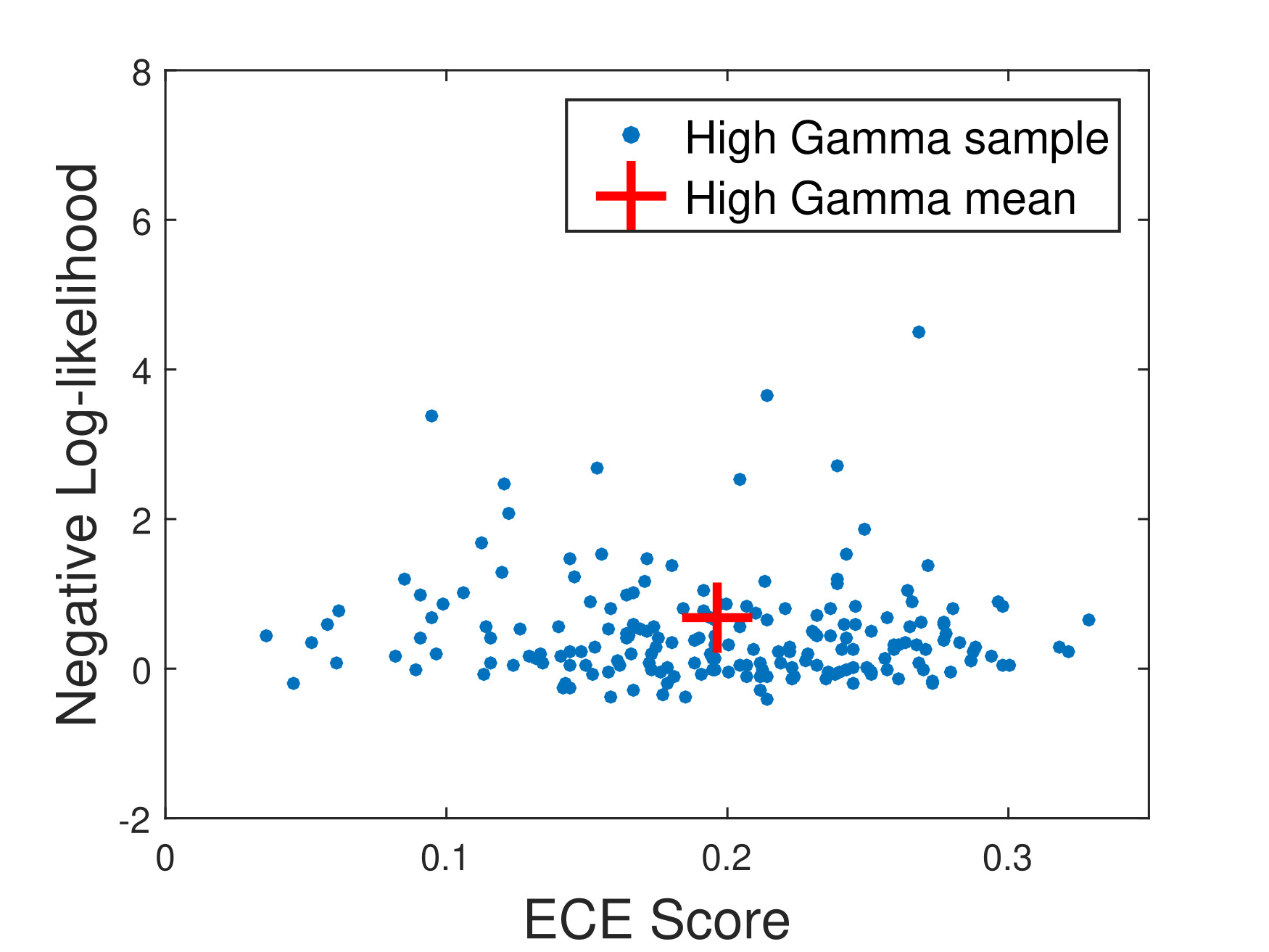}}
  \caption{The correlation between calibration and output uncertainty. 
The relationship of NLL and Brier score are revealed in fig.\ref{7a} and fig.\ref{7c} for BCI-IV 2a and high gamma dataset respectively. Fig.\ref{7b} and fig.\ref{7d} shows correlation between NLL and ECE score. It shows NLL and Brier score have a positive correlation in general while NLL and ECE score are largely uncorrelated.}
  \label{fig7} 
\end{figure}

It is worth analyzing on the relationship between the different uncertainty and calibration metrics. We empirically analyzed the correlation between the two types of calibration error and negative log-likelihood (NLL).
The relationship betweeen NLL and Brier score are revealed in fig. \ref{7a} and fig. \ref{7c} for BCI-IV 2a and high gamma dataset respectively. The two metrics sees a positive correlation in general. Intuitively, lower Brier score indicates predicted $y$ and ground truth $\bar{y}$ being closer to each other, contributing to NLL decrease. Fig. \ref{7b} and fig. \ref{7d} shows NLL and ECE score are largely uncorrelated.

The input noise of EEG source signal can be estimated based on signal to noise ratio (SNR) of the BCI system. Here we adopt a more accurate approach and estimate it by performing a grid search on a range of noise values, as previous study \cite{Loquercio2020AGF} proved NLL is minimized when the magnitude of assumed input noise matches the underlying ground truth. Result on the grid research for BCI-IV 2a and high gamma dataset is provided in Table \ref{t6} and Table \ref{t7} respectively. The input noise of BCI-IV 2a is estimated to have a magnitude of 0.1, and high gamma dataset is endowed with smaller noise magnitude of 0.01. 

\subsection{Uncertainty Reduction Analysis}

\begin{table*}[htbp]
  \centering
  \caption{Summarization on applied corruptions in data augmentation for uncertainty reduction}
    \begin{tabular}{lr}
    \toprule
    Corruption Type & Description \\
    \midrule
    Gaussian noise & adding gaussian signal noise \\
    Shot noise & electronic noise originates from the discrete nature of digital logic units \\
    Impulse noise  & instantaneous sharp discruption like clicks and pops in signal \\
    Motion blur & disruption in signal due to movement of recording electrodes \\
    Zoom blur & disruption in scaling of signal \\
    Intensity & the magnitude variation in EEG signal \\
    Contrast & the fluctuation in ratio of highs and lows in EEG signal \\
    Elastic & transformations stretch or contract among small portions of EEG signal \\
    \bottomrule
    \end{tabular}%
  \label{corruption-types}%
\end{table*}%

We compared the proposed UNCER model to other data augmentation approaches for uncertainty reduction and improvement on model robustness. RandAugment \cite{NEURIPS2020_d85b63ef} is a simple and effective data augmentation method with reduced search space design, MixUp\cite{zhang2018mixup} performs linear interpolation on training examples together with their labels,  MaxUp \cite{maxup} performs data augmentation by optimizing the mixing weights of Mixup in the worst-case. Augmix \cite{hendrycks*2020augmix} composes and combines different augmentation operations with different depths and widths to generate complex corruptions.

The model performance are compared for both intra-subject and cross-subject scenarios. Comparison on BCI-IV 2a dataset is summarized in Table \ref{bci-iv-reduction}. The proposed model significantly reduces uncertainty and calibration error, has an improvement of 38.9\% in terms of NLL, and reduced the Brier error and ECE error by 10.3\% and 9.1\% respectively comparing to best performing counterparts. The proposed method is designed to preserve the semantic information of the original signal data, allowing it to achieve better accuracy and reduced corruption error, e.g. UNCER has a margin of at least 2.4\% on accuracy improvement and 5.0\% on corruption error reduction for intra-subject classification. A margin of 3.3\% on accuracy and 7.2\% on corruption error is observed for the cross subject scenario.

Comparison on high gamma dataset is shown in Table \ref{high-gamma-red}. For intra-subject classification, the proposed method outperforms other comparison methods by at least 13.3\% on NLL, 11.6\% on Brier error and 6.2\% on ECE error. Similar improvements are observed for cross subject classification, where the proposed method achieved an improvement of 26.1\% on NLL, 5.2\% on Brier score and 2.4\% on ECE error. We observed the improvement is more significant on BCI-IV 2a dataset than high gamma dataset, which is related to the fact that high gamma dataset is less noisy and models are having a relatively high prediction accuracy with low variance and calibration error.

\begin{table}[htbp]
  \centering
  \caption{Estimation of BCI-IV 2a input noise level through grid search on NLL}
  \resizebox{0.8\columnwidth}{!}{%
    \begin{tabular}{l|rrrrr}
    \toprule
    Estimated Input Noise & 0.02  & 0.05  & 0.1   & 0.2 & 0.3 \\
    \midrule
    NLL & 6.4   & 6.19  & \textbf{6.11}  & 6.31 & 7.64\\
    \bottomrule
    \end{tabular}%
    }
  \label{t6}%
\end{table}%

\begin{table}[htbp]
  \centering
  \caption{Estimation of high gamma dataset input noise level through grid search on NLL}
  \resizebox{0.8\columnwidth}{!}{%
    \begin{tabular}{l|rrrrr}
    \toprule
    Estimated Input Noise & 0.002 & 0.005 & 0.01  & 0.02 & 0.03\\
    \midrule
    NLL & 12.8  & 2.06  & \textbf{1.78} & 6.47 & 16.5\\
    \bottomrule
    \end{tabular}%
    }
  \label{t7}%
\end{table}%

\subsubsection{Effectiveness of Meta Learning Formulation in Uncertainty Reduction} We formulate the uncertainty reduction process in Algorithm \ref{alg:uncer} as a meta learning procedure. Here we compare it to another alternative process of simultaneously optimizing on the overall model without splitting the operation set into seen and unseen operations. The experiment is conducted on BCI-IV 2a dataset and result is shown in Table \ref{joint-ablation}. For the intra-subject scenario, we observed 21.5\%, 8.24\% and 5.33\% improvement on NLL, Brier score and ECE score respectively. Gains for cross-subject scenario is also significant, with 11.5\% 3.54\% and 2.75\% improvement on NLL, Brier score and ECE score respectively. This shows the effectiveness of meta learning approach in optimizing the augmentation operations.

\subsubsection{Influence of Consistency Loss Hyper Parameter $\lambda$}
Hyper parameter $\lambda$ controls the magnitude of consistency loss term $JS(\widehat{\vx}_{b}, \vx_{orig})$. We performed analysis on the model's sensitivity towards $\lambda$ in Table \ref{ablation-lambda}. We observed improved uncertainty as $\lambda$ increases, and it gradually converges with $\lambda > 10$. This shows that consistency between original and augmented signal is benificial towards model performance.

\begin{table*}[htbp]
  \centering
  \caption{Ablation study on influence of different training approaches towards model performance for BCI-IV 2a dataset}
  \scalebox{0.8}{
    \begin{tabular}{c|l|r|rr}
    \toprule
    \multirow{2}[4]{*}{Scenario} & \multicolumn{1}{c|}{\multirow{2}[4]{*}{Method}} & \multicolumn{1}{l|}{Variance Estimation} & \multicolumn{2}{c}{Calibration} \\
\cmidrule{3-5}          &       & \multicolumn{1}{l|}{NLL} & \multicolumn{1}{l}{Brier} & \multicolumn{1}{l}{ECE} \\
    \midrule
    \multirow{2}[2]{*}{Intra Subject} & UNCER-Joint Learning & 3.76  & 0.085 & 0.244 \\
          & UNCER-Meta Learning & \textbf{2.95}  & \textbf{0.078} & \textbf{0.231} \\
    \midrule
    \multirow{2}[2]{*}{Cross Subject} & UNCER-Joint Learning & 23.4  & 0.113 & 0.255 \\
          & UNCER-Meta Learning & \textbf{20.7}  & \textbf{0.109} & \textbf{0.248} \\
    \bottomrule
    \end{tabular}%
    }
  \label{joint-ablation}%
\end{table*}%

\begin{table}[htbp]
  \centering
  \caption{Ablation study on influence of consistency loss term $\lambda$ towards UNCER model performance for BCI-IV 2a dataset}
    \begin{tabular}{l|rrrrr}
    \toprule
    Value of $\lambda$ & 1     & 5     & 10    & 15 & 20 \\
    \midrule
    NLL   & 4.31  & 3.68  & 3.12  & 2.95 & 2.89\\
    \bottomrule
    \end{tabular}%
  \label{ablation-lambda}%
\end{table}%

\section{Conclusion} 

In this work, we proposed an effective method to accurately estimate and reduce the uncertainty lying in neural decoding of EEG signal. The proposed method considers both the noise from input electrodes and also the randomness lying in model parameters, allowing accurate modeling of the uncertainty in decoding decision. The uncertainty reduction is achieved with an adaptive data augmentation based approach. We performed detailed experiment on two motor imagery classification tasks, where it significantly outperforms current state of the arts in terms of uncertainty estimation and its reduction, and at the same time maintains predictive accuracy on par with the other models. The method can be readily integrated into existing BCI systems for reliability improvement.

\bibliographystyle{IEEEtran}
\bibliography{ue}

\end{document}

%% file: math_commands.tex
\usepackage{amsmath,amsfonts,bm}

\def\eqref#1{equation~\ref{#1}}

\def\1{\bm{1}}

\def\vtheta{{\bm{\theta}}}

\def\vc{{\bm{c}}}

\def\vh{{\bm{h}}}

\def\vo{{\bm{o}}}

\def\vw{{\bm{w}}}
\def\vx{{\bm{x}}}

\def\mI{{\bm{I}}}

\DeclareMathAlphabet{\mathsfit}{\encodingdefault}{\sfdefault}{m}{sl}
\SetMathAlphabet{\mathsfit}{bold}{\encodingdefault}{\sfdefault}{bx}{n}

\def\gL{{\mathcal{L}}}

\def\gO{{\mathcal{O}}}

\def\gS{{\mathcal{S}}}
\def\gT{{\mathcal{T}}}
\def\gU{{\mathcal{U}}}

